\documentclass{pj}
\usepackage{graphicx}
\usepackage{algorithm}
\usepackage[noend]{algpseudocode}
\usepackage{amsmath}

\DeclareRobustCommand{\textsupsub}[2]{{%
  \m@th\ensuremath{%
    ^{\mbox{\fontsize\sf@size\z@#1}}%
    _{\mbox{\fontsize\sf@size\z@#2}}%
  }%
}}

\begin{document}
\setcounter{page}{1}
\pjheader{Vol.\ x, y--z, 2014}

\title[Rapid Designs of wide-area metasurfaces]
{Rapid Design of Wide-Area Heterogeneous Electromagnetic Metasurfaces beyond the Unit-Cell Approximation}
\footnote{\it Received date}  
\footnote{\hskip-0.12in*\, Corresponding author:~Ravi~Hegde (hegder@iitgn.ac.in).}
\footnote{\hskip-0.12in\textsuperscript{1} Department of Electrical Engineering,
Indian Institute of Technology-Gandhinagar, Gandhinagar, Gujarat, India 382355.}

\author{Krupali~D. Donda \textsuperscript{1} and Ravi~S. Hegde \textsuperscript{1,*}}

\runningauthor{Donda and Hegde}

\tocauthor{Krupali ~D. Donda and Ravi~S. Hegde}

\begin{abstract} 
  We propose a novel numerical approach for the optimal design of wide-area
  heterogeneous electromagnetic metasurfaces beyond the conventionally used
  “unit-cell” approximation. The proposed method exploits the combination of
  Rigorous Coupled Wave Analysis (RCWA) and global optimization techniques (two
  evolutionary algorithms namely the Genetic Algorithm (GA) and a modified form
  of the Artificial Bee Colony (ABC with memetic search phase method) are
  considered). As a specific example, we consider the design of beam deflectors
  using all-dielectric nanoantennae for operation in the visible wavelength
  region; beam deflectors can serve as building blocks for other more
  complicated devices like metalenses. Compared to previous reports using local
  optimization approaches our approach improves device efficiency; transmission
  efficiency is especially improved for wide deflection angle beam deflectors.  The ABC method with memetic search phase is
 also an improvement over the more commonly used GA as it reaches similar
 efficiency levels with upto 35\% reduction in computation time.  
 The method described here is of interest
  for the rapid design of a wide variety of electromagnetic metasurfaces
  irrespective of their operational wavelength.
\end{abstract}

\setlength {\abovedisplayskip} {6pt plus 3.0pt minus 4.0pt}
\setlength {\belowdisplayskip} {6pt plus 3.0pt minus 4.0pt}

\section{Introduction}\label{sec:intro}

The
metasurface~\cite{Holloway2012,Genevet2015,Kildishev2012,Lin2014,Meinzer2014,Minovich2015}
is the two-dimensional analogue of the metamaterial.  It is a spatially
heterogeneous array of nanoscale resonant elements (called meta-atoms) that can,
in general, alter the amplitude, phase, spectrum and polarization values of an
incident wave-front in a very short propagation distance and with sub-wavelength
resolution in the transverse plane~\cite{Minovich2015}.  Being a structured
surface, it can be fabricated easily in relation to the metamaterial.
Additionally, it reduces the insertion losses and provides compactness. 

The metasurface is a frequency agnostic concept and is finding applications
across the electromagnetic spectrum in different tasks.  In the optical and
infrared frequency regions, they are of interest in connection with integrated
photonics. Although this concept~\cite{yu2011,yu2014b} has been initially
explored in connection with plasmonic nanoantenna~\cite{Kildishev2012} attention
has now turned towards all-dielectric metasurfaces~\cite{khurgin2015,aieta2012,
decker2014,decker2014a,staude2013}. Such high refractive index dielectric
nanoantenna arrays can also be considered as two dimensional high-index contrast
subwavelength diffraction gratings; various optical wavefront manipulation
possibilities have been demonstrated with these so called hcta (high contrast
transmit arrays)~\cite{arbabi2014b,arbabi2014d,arbabi2016,arbabi2015}.  In the
microwave region also low-profile transmitarrays are of interest for a variety of
applications~\cite{Pfeiffer2013}. Metasurfaces can achieve phase and
polarization control simultaneously and are of interest in achieving millimeter
wave beam-shaping lenses~\cite{Pfeiffer2015,Pfeiffer2013} and
carpet-cloaks~\cite{Hsu2015}. 

The constituent elements of a metasurface, the meta-atoms, are subwavelength
resonators while the transverse extent of the metasurface can be several orders
of magnitude larger than the operating wavelength (in other words, useful
metasurfaces will be electrical large in the transverse plane). For heterogeneous
metasurfaces, this means that the number of free parameters in the design of the
metasurface can exceed $10^9$~\cite{Byrnes2016a}. This structure is not easily
amenable for analysis, synthesis and optimization tasks.  Conventionally, the
so-called unit-cell approximation~\cite{Zhou} has been
adopted~\cite{Byrnes2016a} in the design of heterogeneous metasurfaces, whereby
each constituent meta-atom is designed as if it was part of an infinite periodic
lattice. This unit-cell approach has also been extensively used in reflectarray
design in the microwave domain and it has been reported that it leads to
discrepancies when experiments are compared with the simulated
designs~\cite{Zhou}. The unit cell approximation has several limitations,
chiefly, it tends to be inadequate in the presence of strong phase gradients,
interactions between neighboring pillars, and oblique incidence angles. These
constraints become important for instance in the case of a focusing
lens~\cite{Pfeiffer2015,Pfeiffer2013,Byrnes2016a,Yu2015a}.

In this paper, we propose an approach whereby spatial order is applied on an
extended cell, which includes several individual resonator elements. This
technique is similar to the Extended Local Periodicity (ELP)
approach~\cite{Zhou} proposed for reflectarray design.  Specifically, we focus
our attention to a beam deflector element which is a commonly used structure
for benchmarking metasurfaces design strategies
\cite{Yu2015a,Byrnes2016a,Aviv2017} as the transmission efficiencies can be
compared across designs. The beam deflector element changes the direction of an
normally incident plane wave by giving it a predesigned inclination with respect
to the normal. Although a simple element, the beam deflector, in addition to
being an important element in itself, can be combined to produced metasurfaces
with more advanced functionality like high numerical aperture and
multi-wavelength focusing lenses and holograms
\cite{Byrnes2016a,Aviv2017,Cheng2014,Guo2017}. The beam deflector elements
considered in this paper do not use the local phase approach~\cite{Aviv2017}
which is based on the unit-cell approximation but move beyond this by
considering an extended unit-cell consisting of a larger number of nanoantenna.

The main contribution of this article is that it presents a systematic
investigation of global optimization methods for the design of the extended
unit-cells. In comparison to the beam deflector synthesis method reported by
Byrnes et.\ al.~\cite{Byrnes2016a} which use local optimization methods, our
method which relies on global optimization methods like the Genetic Algorithm and
Artificial Bee Colony show significant improvement in efficiency by not getting
stuch at local optima.
In comparison to the \cite{Aviv2017} which has studied the application of
Genetic Algorithm (GA) to the beam deflector design problem, we have explored
unit-cells which do not restrict the design to a rectangular lattice and
cylindrical elements. Additionally, our Artificial Bee Colony method is seen to
outperform GA method in speed by cutting the convergence time by 35\%. 

The paper is organized as follows: following this introduction, in
section~\ref{sec:method}, we describe the beam deflector geometry and the
optimization algorithms in detail. In section~\ref{sec:gaandmabc}, we discuss
how to phrase the optimization problem in terms of the GA and Memetic ABC
algorithms and suggest how to best choose the hyperparameters of these
algorithms. The performance figures for the designs and comparisons with
previously published reports are presented in section~\ref{sec:resanddis} before
concluding the paper in section~\ref{sec:conclude}.

\section{Specification of the optimization problem} \label{sec:method}

\begin{figure}[h]
\center
\includegraphics[width=1\textwidth,draft=false]{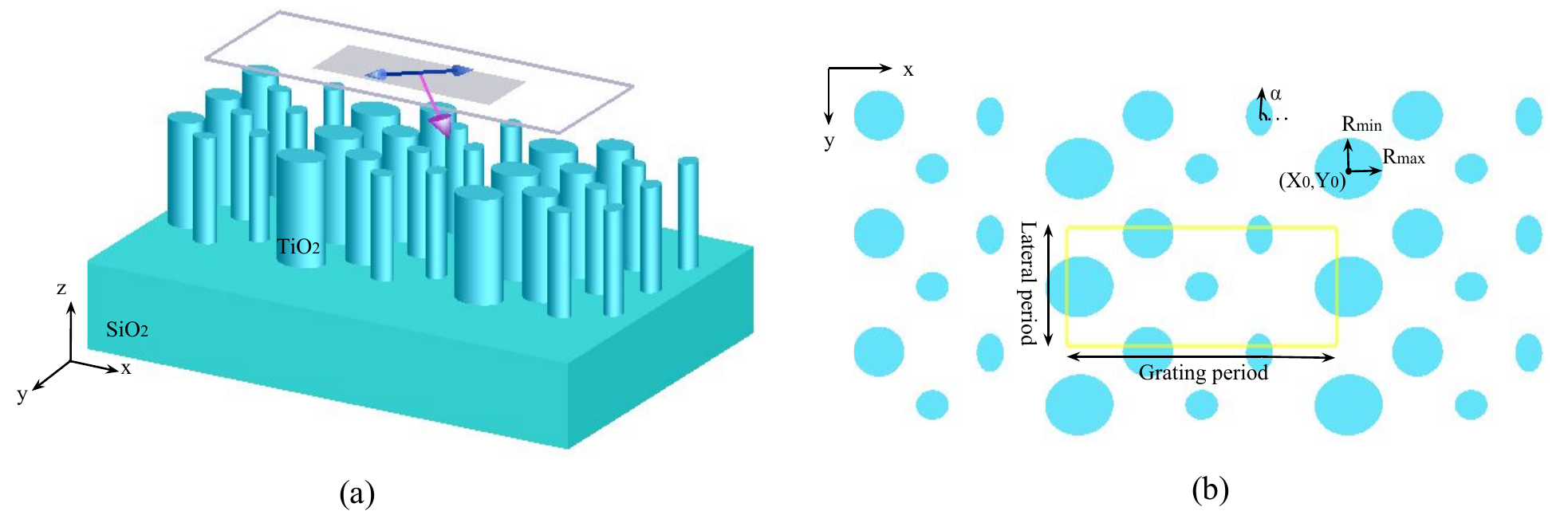}
\caption{(a) Three dimensional schematic of a beam deflector made of extended unit-cells
  containing TiO\textsubscript{2} nano elliptical
antennae on fused silica substrate (b) Design parameters of the extended
unit-cell. The yellow line shows the boundaries of the extended unit-cell.} 
\label{fig:geometry}
\end{figure}

As discussed earlier, the beam deflector geometry is a canonical element in the
design of a wide variety of metasurfaces. The system considered here 
can be directly compared with the geometry
reported in Byrnes et.\ al ~\cite{Byrnes2016a} where a wide-area focusing lens was
designed using beam-deflectors as motifs. Specifically, this required beam
deflectors with deflection angles varying in the range 20 to 70 degrees. 
Figure~\ref{fig:geometry} shows
the geometry of such a beam-deflector with rectangular shaped extended
unit-cells. Here TiO\textsubscript{2}
nanoantennae are arranged in a hexagonal lattice within the unit-cells on a fused silica substrate.
We have kept the overall dimensions of the extended unit-cells for any
particular deflection angle exactly equal to those in Byrnes et.\
al~\cite{Byrnes2016a} so that our designs can serve as a drop-in replacements
with higher efficiencies.

A plane wave is normally incident from the substrate side and upon leaving the
beam deflector emerges at an angle ($\theta$ in the $x-z$ plane) that is decided by the design (vice-versa a
light wave incident at this angle from the free-space side emerges at normal
incidence). The period of the beam deflector unit-cell in the $x$ direction  is called the grating
period (G. P.) and the period in the $y$ direction is called the lateral
period (L. P.).  These periods are given by:
\begin{eqnarray}
    G.P.=\lambda_{0}/sin \theta, \\ \nonumber 
    L.P.= \Phi f \tan{\theta},
    \label{eqn:gplp}
\end{eqnarray}
where, $\lambda_{0}$ is free space wavelength of source light and $\theta$ is
angle of incidence or the desired deflection angle.  $\Phi$ is angular width of
unit cell in $y$ direction and is decided by the lateral width and the desired
focal length. Throughout the paper we consider the lateral period  as 400 nm as in \cite{Byrnes2016a}.
As seen in equation~\ref{eqn:gplp}, a beam deflector with a larger deflection
angle requires a larger period. 

Within each extended unit-cell, we consider elliptically shaped nanoantennae
each with freely choosable values of the following 5 parameters: major and minor
axes  $r_{max}$ and $r_{min}$ respectively,
the $x$ and $y$ coordinate of the center of the ellipse in the cell coordinate
system $x_0$ and $y_0$ respectively, and the orientation angle of the major
axis $\alpha$. In each unit cell the number of
nanoantennae $N$ can be calculated by:
\begin{equation}
    N \le G.P./d_{min},
    \label{eqn:Ni}
\end{equation}
where $d_{min}$ is minimum diameter of nano structure and can be decided based
on the fabrication constraints. Here we impose the condition that the minimum
feature size of the entire shape should not be lower than 100 nm.  

The optimization problem can now be stated as:

\begin{equation}
\begin{aligned}
& \underset{s}{\text{maximize}}
& & 0.5(\eta^1(s) + \eta^2(s)) \\
& \text{subject to}
& & f_i(s) \leq b_i, \; i = 1, \ldots, m,
\end{aligned}
\end{equation}

where $s$ is a $5N$ sized vector describing the extended unit cell ( i.e.\ $s =
[r_{max}^i, r_{min}^i, x_0^i, y_0^i, \alpha^i]$ and $i = 1 \cdots N$),  
$\eta^1$ and  $\eta^2$ describe the first order beam efficiency of the designed
grating for $x$ and $y$ polarized normally incident beams respectively (the
first order is made to coincide with the desired angle $\theta$), and the
various functions $f_i$ describe constraints imposed on the values that the
vector $s$ or its elements can take (for instance, these constraints could
require that the ratio of the major and the minor axes be limited to 5). 
More generally, fabrication techniques impose several constraints on realizable
designs and these could be incorporated as shown above. 
In order to calculate the first order efficiency of a particular extended
unit-cell, we use the rigorous coupled wave analysis (RCWA) approach
\cite{Liu2012} using the open-source S\textsuperscript{4} application.  S\textsuperscript{4} combines the
S-matrix approach with the RCWA method. Full-wave simulations are performed on
the finally obtained optimal vectors to verify the obtained efficiency figures.

\section{Optimization via Genetic Algorithm and Me-ABC Algorithm}
\label{sec:gaandmabc}

Due to the relatively large number of free parameters and stringent fabrication related constraints
required to design efficient metasurface scanning over the full parameter space is not feasible. Stochastic
optimization methods are more appropriate to improve the efficiency and are expected to perform better
than simpler gradient descent based approaches. In order to optimize the polarization averaged first order
diffraction efficiency we employed two techniques: Genetic algorithm and hybrid ABC algorithm with memetic search phase.

\subsection{Genetic Algorithm}
Genetic Search is a very popular method used in electromagnetic problems having
a large multidimensional unknown search space~\cite{Johnson1997,Tremblay2005}.  We have employed genetic algorithm with three common parameters:
selection, crossover and  mutation. Additionally, we have set upper and lower
bounds for the  radii and position of nanoantennae in the beam deflector. Radius
values of ellipses are limited between 50 nm and 125 nm and the minimum distance between
two antennae centres is set to 100 nm. 

\begin{figure}[h]
\centerline{\includegraphics[width=1\columnwidth,draft=false]{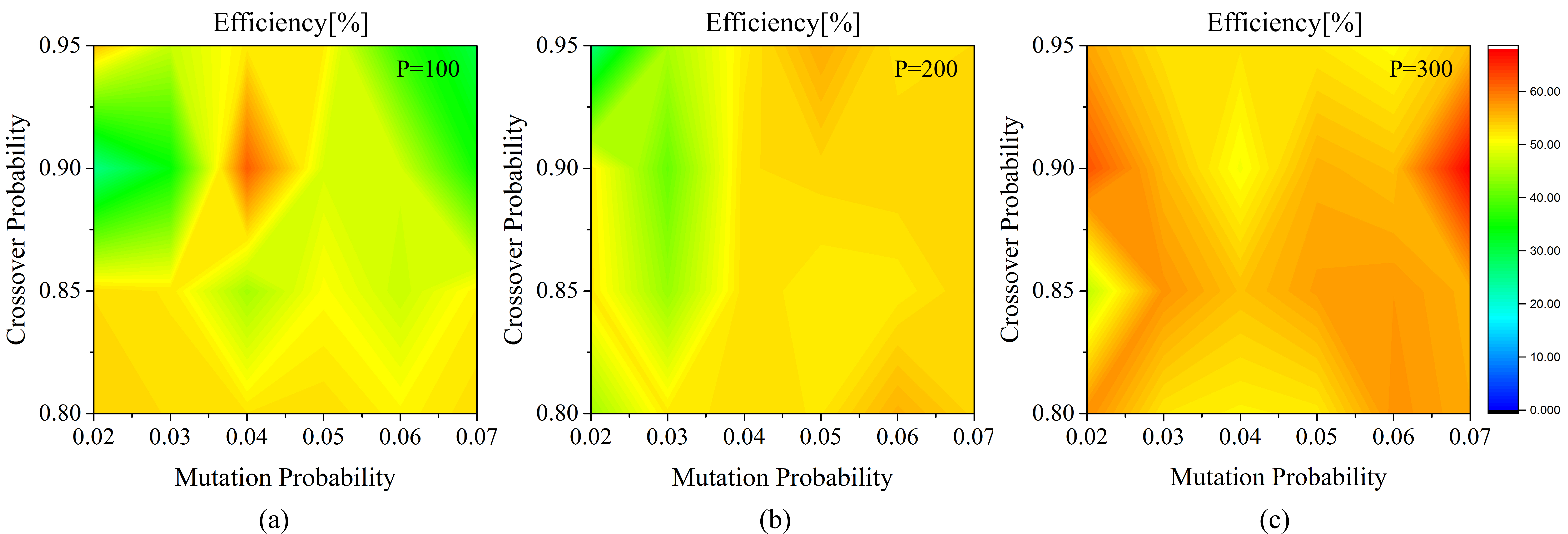}}
\caption{Selection of hyperparameters for properly tuning of GA 
algorithm via grid-search.  (a-c) diffraction efficiency for 70 degree beam
deflector as a function of mutation and crossover
probability for population sizes of 100 (a), 200 (b) and 300 (c).}
\label{fig:gahyper}
\end{figure}

The elliptical nanoantennae represent individuals in a generation.  Each
generation has some number of families $N_f$ represented by beam deflectors with 
$N_i$ number of individuals.  The number of individuals in a family is given by
equation~\ref{eqn:Ni}. The evolution usually starts from a population of
randomly generated individuals that constitute potential metasurface designs.
We are targeting to maximize the polarization averaged first order diffraction
efficiency. For each generation, we choose a fraction $f_t$ of the best performing beam
deflectors and a fraction $f_d$ of individuals randomly out of remaining $1 -
f_t$ fraction of the  lesser
performers from the population based on diffraction efficiency. These
individuals then act as parents for next generation so that we promote genetic
diversity in addition to fitness.  The Crossover step applied next is the
process of taking more than one parent solutions and producing a child solution
from them. The crossover operator given by $P_{c}$ modifies the children
generation by randomly mixing couples. After crossover, We mutate radii,
position and angle of  nano-ellipses by some amount given by mutation
probability $P_{m}$.  The pseudocode for the GA used here is shown below. 

\begin{algorithm}
\caption{Genetic Search Method}\label{alg:ga}
\begin{algorithmic}[1]
\State Generate initial population of size $N_{f}$;
\State Evaluate initial population according to fitness criteria;

\While{Termination criteria meets}
\State Select the best fit individuals for reproduction;
\State Perform crossover operation given by probability $P_{c}$;
\State Perform mutation operation given by probability $P_{m}$;
\State Evaluate fitness of new individuals;
\State Replace the worst individuals of population by best new individuals;
\EndWhile

\State Report the best family achieved
\end{algorithmic}
\end{algorithm}

We have used traditional Grid search method for tuning of hyper parameters of
the GA such as $f_t, f_d, P_c, P_m$ and $N_f$.  It was found that values of 0.15
and 0.10 was best suited for $f_t$ and $f_d$ respectively.  The procedure
followed in the grid search method is demonstrated in figure~\ref{fig:gahyper}.
It shows how different population size, mutation rates and crossover rates affect
the convergence of the efficiency function.  Population diversity is crucial to
the genetic algorithm’s ability to continue fruitful exploration without getting
stuck at local maxima. The size of the population $N_f$ dictates the available
diversity. While a large value will lead to good diversity it will also increase
the computational time. We observe that 300 families in each generation is
sufficient to produce good fitness. The crossover rate $P_{c}$ and mutation rate
$P_{m}$ are both continuous variables within [0,1]. These two values are
important for controlling the balance between exploration and exploitation.
Table~\ref{tab:hyperpars} summarizes a good set of values of all the
hyperparameters for the GA used in this problem.

\subsection{ABC algorithm with Memetic Search Phase}

 Artificial bee colony (ABC) optimization algorithm is a relatively new population based probabilistic
approach for global optimization based on the concept of swarm intelligence \cite{Karaboga2007}. ABC has outperformed many
 other nature inspired optimization algorithms prompting our interest in it for
 metasurface design.  Memetic algorithms (MA) is another growing area of
 research in evolutionary computation. It is inspired by Darwin's principle of
 evolution and Dawkin's notation of a "meme" ~\cite{Moscato2003}. The ABC
 algorithm has achieved excellent results when solving continuous and
 combinatorial optimization problems ~\cite{Bolaji2013}. To achieve the benefits
 of both algorithm, ABC is hybridized with Memetic Algorithm ~\cite{Fister2012,Bansal2013,Kumar2014}. 

In the ABC algorithm ~\cite{Karaboga2007}, the search of a parameter space is
 accomplished by a set of honey bees called the swarm containing three types of
 bees: employed bees, onlooker bees, and scout bees.  Consider an objective
 function $f_{i}(s)$ which evaluates some performance metric of an extended
 unit-cell represented by $s = [r_{max}^i,r_{min}^i, x_0^i, y_0^i, \alpha^i]$
 (here $s$ is a vector of $5N$ geometrical parameters and $N$ is number of
 ellipses on a beam deflector). Each beam deflector is represented by a food
 source $s$ in the swarm and it is generated as follows \cite{Fister2012}:

\begin{equation}          
s_{ij}= s_{minj}+rand[0,1] (s_{maxj}- s_{minj}),
\label{eqn:foodsource}
\end{equation}

where function rand[0,1] returns a random value between 0 and 1, $s_{maxj}$ and
$s_{minj}$  represents maximum and minimum limit of the candidate solution
$s_{i}$. The swarm will now navigate this parameter space
 and converge to an optimal geometry. The parameter space is considered a region
 of space where food sources are located; those regions of this parameter space
 that exhibit a high value of the function $f$ are considered to be richer food
 sources.  The solution search equation of the original ABC algorithm is
 significantly influenced by a random quantity which helps in exploration at the
 expense of exploitation of the search space; there is a significant chance of
 skipping true solutions due to the large step sizes that are often used.  In
 order to balance between diversity and convergence capability of the ABC (in
 other words between exploration and exploitation), a new local search phase is
 integrated with the basic ABC to exploit the search space identified by the
 best individual in the swarm.  The addition of a memetic search phase to the
 ABC algorithm results in the  the memetic ABC (MeABC) algorithm wherein the
 step size required to update the best solution is controlled by a Golden
 Section Search (GSS) approach.  The MeABC has an enhanced exploitation
 capability in comparison to the bare ABC Artificial bee colony algorithm \cite{Fister2012,Bansal2013,Kumar2014}. 
 
 In employed bee step, current solutions are changed by employed bees based on
their individual experience. If the fitness of latest solution is better than
previous solution, then the employed bees update their position to new solution.
The employed bees in ABC algorithm use the following equation in order to
improve self solution \cite{Bansal2013}:
\begin{equation}
v_{ij}= s_{ij} + \Phi_{ij} (s_{ij}-s_{kj}),   k \neq i,	
\label{eqn:e}
\end{equation}
where $v_{ij}$ is the updated food, $s_{i}$ is the old food, $s_{kj}$ is a random
food from hive. Here $k \in{I, 2, ... , s_{N}}$ and  $ j \in {I, 2, ... ,N}$ are
haphazardly chosen indices and $\Phi_{ij}$ is a random number in range $[-1,1]$.
Onlooker bees in the hive expect information of fresh solutions and their position.
In the next step onlooker bees inspect the available information and pick a
solution with a probability given by
\begin{equation} 
P_{i}=\dfrac{fit_{i}}{\sum_{j=1}^{n} fit_{i}},
\label{eqn:p}
\end{equation}
where $fit_{i}$ is $i^{th}$ solution in the swarm. If the position of a food source is not updated for a given cycle it is
considered to be abandoned.  In the scout bee step, the bee whose food source
has been deserted becomes a scout bee and the deserted food source is replaced
by a haphazardly chosen food source within the search space. Scout bees are
agents for global food search; they replace a food source by another randomly chosen
food source which is generated by the equation  
\begin{equation}         
s_{ij}= s_{minj}+rand[0,1](s_{maxj}-s_{minj}), j \in {1,2 \dots N},
\end{equation}
where $s_{minj}$ and $s_{maxj}$ are the bounds of $s_{ij}$ in the
j\textsuperscript{th} direction.

Memetic phase is designed on the golden grid/section search criteria. Basically
we choose a negative value of a (that is some value on the left of the X-axis
for that dimension of the best food) and equal but positive value of b (that is
some value on the right of the X-axis) \cite{Moscato2003}. The values a and b
dictate how wide the memetic search should be. The current best food lies
exactly in the middle initially.  The memetic search starts by determining
whether a better food is on the Left of the current best food (i.e. more towards
a) or to the right (i.e. more towards b). Similar to a binary search algorithm
it updates the values of a and b and gradually zeroes in on the better solution
around the current best. If it finds a better food source, it updates the best
food else the search fails. This process is carried out until the difference
between a and b (absolute value of the difference) is greater than the set value
of epsilon; epsilon being the stopping criteria of memetic search phase. In
MeABC algorithm, ABC algorithm behaves as a local search algorithm in which only
the best individual of the current swarm updates itself in its neighbourhood
while in memetic search phase the step size required to update the best
individual in the current swarm is controlled by the golden section search (GSS)
approach. GSS processes the interval [a=-0.75, b=0.75] and generates two
intermediate points:
\begin{equation}
\begin{aligned}
  F_{1}=b-(b-a) \times \Psi, \\
  F_{2}=a + (b-a) \times \Psi,
\end{aligned}
\end{equation}
where $\Psi$ is the golden ratio. Memetic ABC algorithm has three steps similar to
the ABC algorithm and one more step, the memetic phase, is added for updating
the location of an individual. It changes position~\cite{Bansal2013} given by
the equation
\begin{equation}
  s_{ij}= s_{i} + \Phi_{i}(s_{i}-s_{k})+\Psi(s_{best}-s_{i}),
\end{equation}
where $\Phi_{i}$ is a random number in the interval $[0,D]$ and $D$ is a
positive constant. The pseudo code for the Memetic ABC algorithm that we have
employed is given below. 

\begin{algorithm}
\caption{MeABC Algorithm}\label{alg:meabc}
\begin{algorithmic}[1]
\State Generate an initial population of food sources using equation~\ref{eqn:gplp};
\State Evaluate initial population according to fitness criteria;
\While{Termination criteria meets}
\State Deploy employed bee searches to find new food searches in the neighbourhood using equation~\ref{eqn:e}; 
\State Calculate Probability P for each food source using equation~\ref{eqn:p};
\State Send onlooker bee to food source depending upon P;
\State Evaluate fitness of each new food source;
\If{any employed bee becomes scout bee}
\State Send scout bee to a randomly generated food source;
\EndIf
\State Employ memetic search phase;
\State Memorize the best food source achieved so far;
\EndWhile
\State Report the best food source achieved
\end{algorithmic}
\end{algorithm}

From equation~\ref{eqn:foodsource}, it can be observed
 that the step size consists of a random component $\Phi$ and thus a  proper
 balance is not possible manually~\cite{Bansal2013} ($\Phi$ is random component
 that decides direction and step size of an individual). Memetic search phase
 (MSP) improves the exploitation capability~\cite{Wang2009} considerably.  We
 have used MSP to fine tune the value of $\Phi$ dynamically and iteratively
 using the Golden Section Search strategy. The range of $\Phi_{ij}$ is set to
 $[a,b]$ where $a =-0.75$ and $b = 0.75$; $a$ and $b$ dictate how wide the memetic
 search should be. To tune the hyper parameters of MeABC algorithm like the
 swarm population and numbers of each kind of bees, we used a traditional grid
 search method. The hyperparameter values that were found to yield good results
 in terms of efficiency are summarized in Table~\ref{tab:hyperpars}. 
 
\begin{table}[h]
\renewcommand{\arraystretch}{1.6}
\caption{Hyperparameters chosen for the GA and MeABC method }
\label{tab:hyperpars}
\begin{center}
\begin{tabular}{|p{5cm}|p{1cm}|p{5cm}|p{1cm}|}
		\hline  \multicolumn{2}{|c|}{GA Parameters} & 
		\multicolumn{2}{|c|}{Me-ABC Parameter}  \\
\hline   Parameters & Value & Parameters & Value \\ 
\hline No. of Families  & 300 & No. of Food Sources & 25 \\ 
\hline  No. of Individuals in a Family & 2-4 & No. of Onlooker Bees & 25 \\ 
\hline  Mutation Probability, $P_{r}$ & 0.07 & No. of Employed Bees & 25 \\ 
\hline  Crossover Probability, $P_{c}$ & 0.90  & No. of Scout Bees & 1 \\ 
\hline  No. of Generations & 50 & No. of iterations before bee is tired & 200 \\ 
\hline 
\end{tabular}
\end{center}
\end{table}

\section{Results and discussion} \label{sec:resanddis} 

The above design methodology was applied to the design of a set of beam
deflecting metasurfaces operating at 580 nm wavelength.  The extended unit cells
are rectangular shaped but within it the fill fraction is maximized by adopting
a hexagonal grid. The elliptically shaped nanoantenna are made of $TiO_{2}$ and
sit atop a fused silica substrate. The refractive index of $n(TiO_{2}) = 2.37$
at $580 nm$ and the nanoantenna height is kept at $h=550 nm$.

\begin{figure}[h]
 \centerline{\includegraphics[width=1\columnwidth,draft=false]{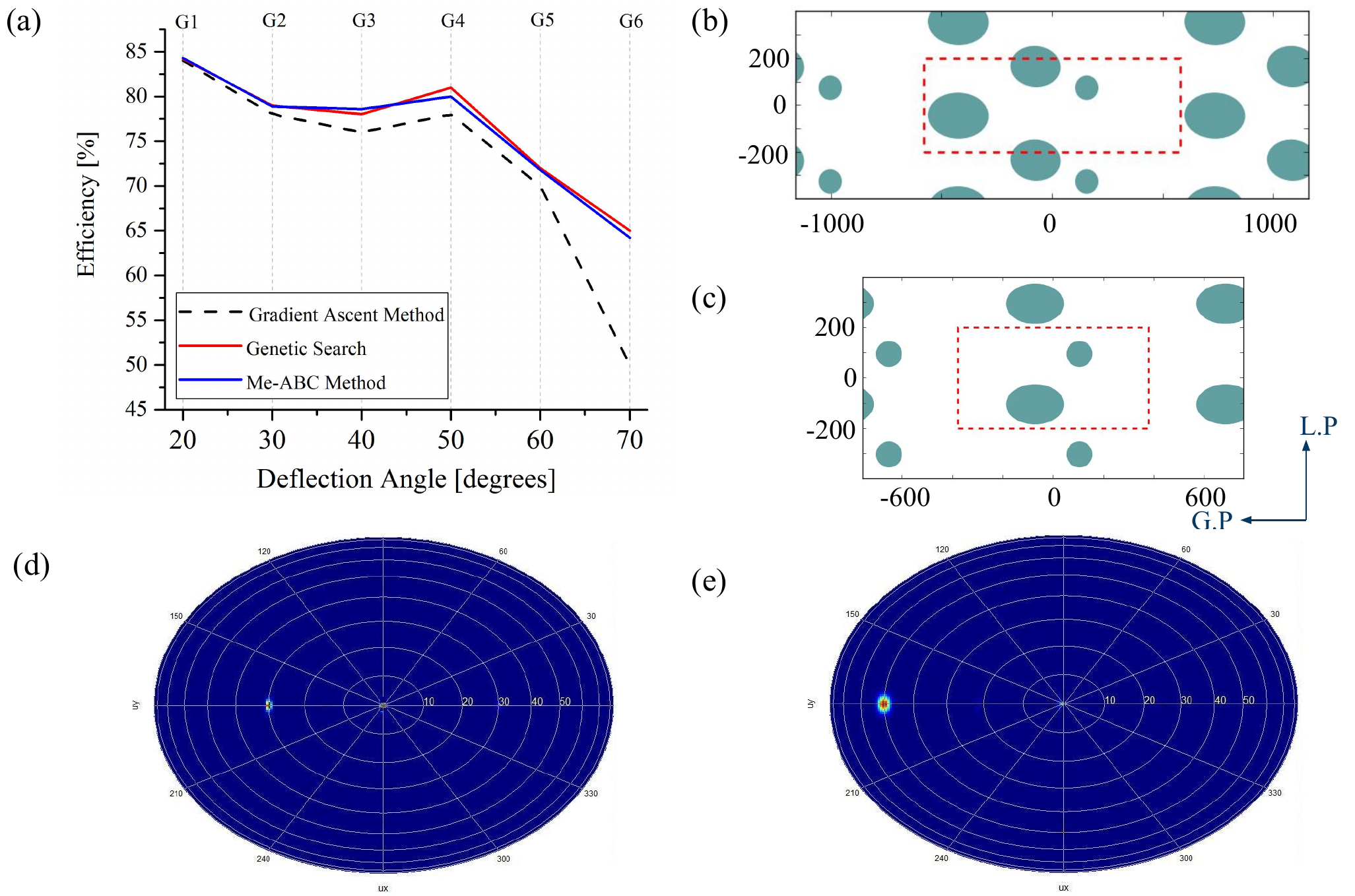}}
 \caption{(a) Comparison of the polarization-averaged first order diffraction efficiencies for beam deflectors with bend angles ranging from 20 to 70 degrees designed via Gradient Ascent, Genetic and
MeABC Approach. Top view of the optimal beam deflectors at deflection angles of 30 degree (b) and 50 degrees (c) (red lines delineate the extended unit-cell). Full-wave simulation based far field intensity plots (d) and (e) for the gratings (b) and (c) respectively.  }
   \label{fig:gratingres}
 \end{figure}

\begin{table}[h]
\renewcommand{\arraystretch}{1.6}
\caption{Optimal geometrical parameters of the extended unit-cells and their
first order polarization-averaged diffraction efficiencies [\%] at $580nm$ wavelength}

\label{tab:finaldesign}
\resizebox{\textwidth}{!}{
\begin{tabular}{|c|c|c|c|c|c|c|c|c|c|c|c|c|c|}
		\hline \multicolumn{1}{|c|}{} &  \multicolumn{3}{|c|}{E1} & 
		\multicolumn{3}{|c|}{E2}  &
		\multicolumn{3}{|c|}{E3}  &
		\multicolumn{3}{|c|}{E4}  &\\
	
\hline D.A. & Rx & Ry & $\theta_{r}$ &  Rx & Ry & $\theta_{r}$  & Rx & Ry & $\theta_{r}$  & Rx & Ry & $\theta_{r}$ & Eff.\\

\hline 20$^{\circ}$ & 69.15  & 83.17 &-9.6 & 50 & 50 & 0&50&50&0&162.2&89.14&0.01&83.8\% \\ 
		
\hline 30$^{\circ}$ &111.14 & 85.63 &175.7&      50.04 & 50.06 & 175.16&     95.99&134.80&88.66&-&-&-&79\% \\

\hline 40$^{\circ}$& 93.25 & 101.76&78.36&50 & 50 & 0&-&-&-&-&-&-&78.6\% \\ 
\hline  50$^{\circ}$&58.09 & 77.46& 179.99 & 50 & 50 & 0&-&-&-&-&-&-&81\% \\ 
\hline  60$^{\circ}$& 106.21 & 79.77 & -179.96 & 50 & 50 & 0&-&-&-&-&-&-&71.8\% \\
\hline  70$^{\circ}$&103&89.86&-177.76&50 & 50 & 0&-&-&-&-&-&-&65\% \\ 
\hline
\end{tabular}
}\end{table}

Figure~\ref{fig:gratingres} (a) shows comparison of the polarization averaged
efficiencies obtainable with all three methods viz: the local steepest gradient
ascent~\cite{Byrnes2016a}, the genetic algorithm, and ABC with memetic phase
search. Firstly, note that all three curves indeed follow the well known fact
that efficiencies are nearly constant up to deflection angles of about 50 degrees
but start to decay rapidly afterwards. While all three methods give nearly equal
efficiencies in the initial angular range, the designs obtained with the global
optimization methods significantly outperform at steeper angles.

 \begin{figure}[h]
 \centerline{\includegraphics[width=1\columnwidth,draft=false]{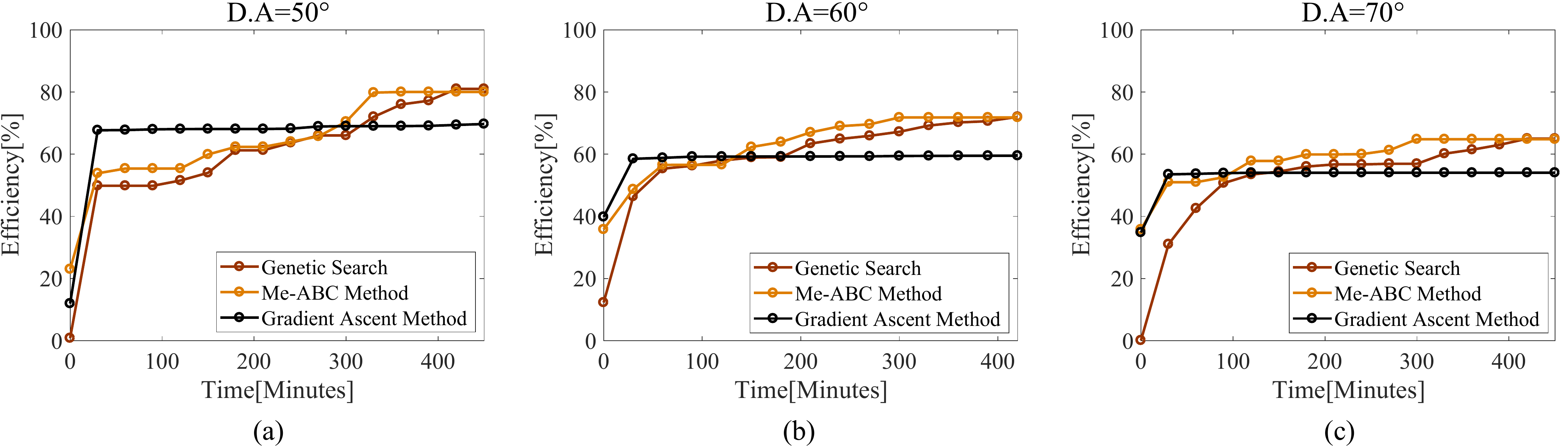}}
 \caption{Comparison of convergence time needed to arrive at the optimal beam deflectors for gradient ascent, GA and MeABC methods. Three different deflection angles have been considered.}
 \label{fig:convergence}
 \end{figure}

Comparison of the convergence times using all three methods for three different
beam deflectors is shown in figure~\ref{fig:convergence}. All runs were made on
machine with the following configurations: Processor - Intel (R) Xeon(R)  CPU
E5-2650 V2@2.60GHz; RAM - 32 GB; and, System Type - Windows 64 bit operating
system.  It is evident that the gradient ascent method gets stuck early on in a
local optima and never improves the overall efficiency. The global optimization
techniques clearly avoid this problem. Furthermore, the time taken by the global
search is not too large in comparison to the local methods. While the
efficiencies achieved by GA and MeABC are almost similar, the MeABC approach
converges to the final geometry in approximately $35\%$ time compared to the GA
approach.




 \section{Conclusion} \label{sec:conclude}
 We have presented two global optimization techniques based on nature inspired
 algorithms for the rapid design of all-dielectric metasurfaces and have applied
 them to the case of beam deflectors.  Compared to local
 optimization methods like the gradient ascent algorithm, Genetic and Me-ABC
 algorithms provides larger deflection efficiencies. Up to $15\%$ efficiency
 improvement is achieved for higher deflection angles. The MeABC method proposed
by us is significantly faster than previously proposed gradient ascent
algorithm~\cite{Byrnes2016a}. It also outperformed the GA in terms of
computation time.   

 \ack
 The authors acknowledge support from the Department of Science and Technology,
 Govt. of India through the Extramural grant SB/S3/EECE/0200/2015.

\end{document}